\documentclass[12pt]{article}
\usepackage[margin=25mm]{geometry}
\usepackage{amsmath,mathtools,amsfonts,bbm}
\usepackage{graphicx,psfrag,epsf}
\usepackage{enumerate}
\usepackage{natbib}
\usepackage{epstopdf}
\usepackage{booktabs}
\usepackage{placeins}
\usepackage{setspace}
\usepackage{float}
\usepackage{soul,color}
\usepackage{placeins}
\usepackage{upgreek}
\usepackage{algorithm,algpseudocode,tabularx}
\usepackage{caption}
\usepackage{subcaption}
\usepackage{enumitem}
\usepackage{multirow}
\usepackage{amssymb,graphics,amsthm, comment}
\usepackage{threeparttable}
\usepackage{url}
\RequirePackage[colorlinks,breaklinks,linkcolor=blue,citecolor=blue,urlcolor=blue]{hyperref}
\usepackage{multicol}
\usepackage{listings} 
\usepackage{xcolor} 
\immediate\write18{texcount -tex -sum  \jobname.tex > \jobname.wordcount.tex}

\graphicspath{{./plots/}}
\newtheorem{theorem}{Theorem}

\newtheorem{condition}{Condition}
\newtheorem{example}{Example}

\newtheorem{corollary}{Corollary}

\allowdisplaybreaks

\providecommand{\keywords}[1]
{
  \small	
  \textbf{\textit{Keywords---}} #1
}

\title{Functional Cox model for interval-censored data}
\author{Yangjianchen Xu $^{1}$,  and Peijun Sang$^{1}$ \\
        \small $^{1}$Department of Statistics and Actuarial Science, University of Waterloo \\
}
\date{} 

\begin{document}
\maketitle

\begin{abstract}
Interval-censored data arise frequently in scientific studies, where the event of interest is known only to occur within a specific time interval. In such studies, functional covariates taking the form of continuous curves or spatial profiles are increasingly encountered, and it is of substantial scientific relevance to investigate how the trajectory of a functional covariate affects the event time. We formulate the effects of both scalar and functional covariates on the interval-censored event time through a functional Cox model. We consider penalized maximum likelihood estimation for this model and devise an EM algorithm to stably compute the parameter estimators. The resulting estimators for the regression parameters and linear functionals of the coefficient function are shown to be consistent and asymptotically normal, with limiting covariance matrices that attain the semiparametric efficiency bound and can be readily estimated through the profile likelihood method. Building upon these results, we construct a global test for the overall effect of the functional covariate. Finally, we assess the performance of the proposed methods through extensive simulation studies and present an application to data from the Alzheimer's Disease Neuroimaging Initiative.
\end{abstract} \hspace{10pt}


\section{Introduction}
Interval-censored data arise when the event of interest is known only to fall within a random time interval induced by periodic examinations. A prominent example is the Alzheimer's Disease Neuroimaging Initiative (ADNI) \citep{adni}, where clinical evaluations are conducted periodically to track the progression of Alzheimer's disease. This large-scale longitudinal study collects both scalar covariates, including demographic and genetic factors, and functional covariates, such as the cortical thickness of the cerebral cortex, which is a continuous function defined over the anatomical surface of the brain \citep{lerch2005focal}. In clinical research, investigators are often interested in evaluating the effects of these covariates on the time to disease progression. In particular, it is of clinical relevance to characterize the effect of the entire trajectory of the functional covariate, which captures its global variations across the continuous domain. However, the fact that none of the event times is exactly observed, together with the infinite-dimensional nature of the functional covariates, makes the regression analysis theoretically and computationally challenging.

Methods for analysing interval-censored data with scalar covariates have been extensively studied \citep[e.g.,][]{Huang1996, Huang&Wellner1997, Cai2003hazard, Gu2005, zhang2005regression, zhang2013empirical, Zeng2016}. When functional covariates are incorporated, existing regression models for time-to-event outcomes focus exclusively on right-censored data. Specifically, \citet{Chen2011string} proposed a functional Cox regression model for right-censored data. \citet{Lee2015bflcrm} and \citet{kong2018flcrm} studied this model under a Bayesian framework and through functional principal component analysis, respectively. \citet{qu2016optimal} and \citet{hao2021semiparametric} further investigated the estimation of this model under the reproducing kernel Hilbert space framework and established the asymptotic properties of the parameter estimators. Beyond the Cox model, \citet{Jiang2020functional} and \citet{liu2025efficient} developed a functional quantile regression model and a functional accelerated failure time model, respectively, to study the effects of functional covariates on right-censored event times. \citet{guo2026semiparametric} developed a joint modelling approach to characterize the association between the right-censored event time and a concurrent functional covariate subject to measurement error.

There is no existing method for regression analysis of interval-censored data with functional covariates. This task is challenging for several reasons. First, analysing interval-censored data is more difficult than analysing right-censored data. Interval-censored data provide only the time intervals that bracket the unobserved event times, yielding a more complex likelihood function and rendering standard techniques for right-censored data, such as the partial likelihood principle \citep{partial} and the counting process framework \citep{Andersen&Gill}, inapplicable. Second, incorporating functional covariates further increases the computational complexity. The effect of a functional covariate is typically characterized by an infinite-dimensional coefficient function; estimating this function nonparametrically requires sophisticated numerical methods. Finally, establishing the asymptotic properties of the estimator for the coefficient function under interval censoring is highly nontrivial and requires novel adaptations of empirical process theory \citep{van1996weak} and semiparametric efficiency theory \citep{bickel1993efficient}.

In this article, we study the functional Cox model for general interval-censored data with both scalar and functional covariates. We adopt a penalized maximum likelihood framework to estimate the model parameters. By establishing a representer theorem that projects the infinite-dimensional coefficient function onto a finite-dimensional subspace, we devise an EM algorithm in which the estimator for the cumulative baseline hazard function has an explicit form and the estimating equations for the remaining parameters yield unique solutions. Under mild regularity conditions, we establish the consistency and convergence rates of the proposed estimators. Furthermore, we show that the estimators for the regression parameters and linear functionals of the coefficient function are asymptotically normal. Their limiting covariance matrices attain the semiparametric efficiency bound and can be analytically estimated via the profile likelihood method \citep{profile}. Based on these results, we formulate a global test for the overall effect of the functional covariate by deriving a Wald-type statistic.

The remainder of the article is organized as follows. Section \ref{sec:model} introduces the data structure and formulates the functional Cox model for interval-censored data. In Section \ref{sec:estimation}, we exploit a representer theorem to devise an EM algorithm for model fitting. Section \ref{sec:theory} establishes the asymptotic properties of the parameter estimators, which subsequently provide the theoretical foundation for the formal inference procedures developed in Section \ref{sec:infer}. The finite-sample performance of the proposed methods is evaluated through extensive simulation studies in Section \ref{sec:simu}, and illustrated with an application to data from the Alzheimer's Disease Neuroimaging Initiative in Section \ref{sec:application}. Section \ref{sec:conclusion} concludes the article. All technical proofs are relegated to the Supplementary Material.

\section{Data, Model and Likelihood}
\label{sec:model}

Let $T$ denote the event time of interest. Suppose $\boldsymbol{X}(t)$ is a $p$-vector of potentially time-dependent covariates, and $Z(s)$ is a real-valued functional covariate observed over a domain $\mathcal{S}$. The functional Cox model specifies that the hazard function for $T$, conditional on $\boldsymbol{X}$ and $Z$, takes the form
\begin{equation}
	\label{Cox}
	\lambda(t \mid \boldsymbol{X}, Z)=\lambda(t)\exp\left\{\boldsymbol{\alpha}^{\top}\boldsymbol{X}(t)+\int_{\mathcal{S}}\beta(s)Z(s)\mathrm{d}s\right\},
\end{equation}
where $\boldsymbol{\alpha}$ is a $p$-vector of regression parameters, $\beta(s)$ is an unknown coefficient function, and $\lambda(t)$ is an arbitrary baseline hazard function. The cumulative baseline hazard function is defined as $\Lambda(t)=\int_{0}^{t}\lambda(u)\mathrm{d}u$. We denote the full parameter of interest as $\boldsymbol{\theta}=(\boldsymbol{\alpha},\beta,\Lambda)$.

We formulate the interval censoring by assuming that there exists a random sequence of examination times, denoted by $U_{1}<\cdots<U_{M}$, where the number of examination times $M$ is random. Write $\boldsymbol{U}=(U_{0},\ldots,U_{M+1})$, where $U_{0}=0$ and $U_{M+1}=\infty$. We define $\boldsymbol{\Delta}=(\Delta_{0},\ldots,\Delta_{M})$ as the indicators of event status, where $\Delta_{k}=I(U_{k}<T\leq U_{k+1})$ ($k=0,\ldots,M$), and $I(\cdot)$ is the indicator function. The interval-censored data $\{O_i=(\boldsymbol{U}_{i}, \boldsymbol{\Delta}_{i}, \boldsymbol{X}_i, Z_i),i=1,\ldots,n\}$ consist of $n$ independent and identically distributed realizations of $O=(\boldsymbol{U}, \boldsymbol{\Delta}, \boldsymbol{X}, Z)$. We assume that $(\boldsymbol{U},M)$ is independent of $T$ conditional on $\boldsymbol{X}$ and $Z$.

Under model \eqref{Cox}, the loglikelihood function for $\boldsymbol{\theta}$ based on a single observation $O$ is
\begin{align*}
	l(\boldsymbol{\theta})&=\sum_{k=0}^{M}\Delta_{k}\log\left(\exp \left[-\int_0^{U_{k}} \exp\left\{\boldsymbol{\alpha}^{\top}\boldsymbol{X}(t)+\int_{\mathcal{S}}\beta(s)Z(s)\mathrm{d}s\right\} \mathrm{d}\Lambda(t)\right]\right.\\
	&\quad\quad\quad\quad\quad\quad\quad\quad\quad\quad\quad\quad\left.-\exp \left[-\int_0^{U_{k+1}} \exp\left\{\boldsymbol{\alpha}^{\top}\boldsymbol{X}(t)+\int_{\mathcal{S}}\beta(s)Z(s)\mathrm{d}s\right\} \mathrm{d}\Lambda(t)\right]\right),
\end{align*}
where $\int_0^{\infty} \exp\{\boldsymbol{\alpha}^{\top}\boldsymbol{X}(t)+\int_{\mathcal{S}}\beta(s)Z(s)\mathrm{d}s\} \mathrm{d}\Lambda(t)$ is defined as $\infty$. For the parametric component $\boldsymbol{\alpha}$, we denote its parameter space by $\mathcal{A}$, which is a known compact set in $\mathbb{R}^{p}$. The coefficient function $\beta$ is assumed to reside in a reproducing kernel Hilbert space $\mathcal{H}$ equipped with the reproducing kernel $K(\cdot,\cdot)$. By Mercer's theorem, the reproducing kernel admits the spectral decomposition $K(w, s) = \sum_{j=1}^{\infty} \kappa_{j} \phi_{j}(w) \phi_{j}(s)$, where $\{\kappa_{j}\}_{j=1}^{\infty}$ is a nonincreasing sequence of nonnegative eigenvalues, and $\{\phi_{j}\}_{j=1}^{\infty}$ is a set of the corresponding eigenfunctions. For the cumulative baseline hazard function $\Lambda$, the analysis is restricted to the interval $[0,\tau]$, defined as the union of the supports of $U_{k}$ ($k=1, \ldots, M$). The parameter space of $\Lambda$, denoted by $\mathcal{F}$, is the collection of monotonically increasing functions $\Lambda:[0,\tau]\mapsto [0,\infty)$ with $\Lambda(0)=0$. Finally, the total parameter space of $\boldsymbol{\theta}$ is denoted by $\Theta=\mathcal{A}\times\mathcal{H}\times\mathcal{F}$.

\section{Estimation Procedure}
\label{sec:estimation}
\subsection{Penalized Estimator and Representer Theorem}
\label{sec:est}
We estimate the parameter $\boldsymbol{\theta}$ by maximizing the penalized loglikelihood function:
\begin{equation*}
	\widehat{\boldsymbol{\theta}} = (\widehat{\boldsymbol{\alpha}},\widehat{\beta},\widehat{\Lambda}) = \arg\max_{\boldsymbol{\theta}\in \Theta} \left[ \mathbb{P}_{n}\{l(\boldsymbol{\theta})\} - \gamma_{n} J(\beta) \right],
\end{equation*}
where $\mathbb{P}_{n}$ denotes the empirical measure from $n$ independent observations, $J(\cdot)$ is a penalty function controlling the smoothness of the functional coefficient $\beta$, and $\gamma_{n} > 0$ is a tuning parameter that may depend on $n$. Directly solving this maximization problem is challenging because $\beta$ and $\Lambda$ are infinite-dimensional and are entangled with the finite-dimensional $\boldsymbol{\alpha}$ in the loglikelihood function. To address this issue, we show that $\widehat{\beta}$ has a closed-form representation and then develop an EM algorithm for computation.

Let the penalty function $J(\cdot)$ be a squared seminorm on $\mathcal{H}$ such that the null space $\mathcal{H}_{0} = \{\beta \in \mathcal{H}: J(\beta)=0\}$ is a finite-dimensional linear subspace of $\mathcal{H}$. This specific choice of penalty induces an orthogonal decomposition $\mathcal{H} = \mathcal{H}_0 \oplus \mathcal{H}_1$, where $\mathcal{H}_1$ is the orthogonal complement of $\mathcal{H}_0$. Consequently, the reproducing property ensures that the kernel function decomposes as $K = K_{0} + K_{1}$, where $K_{0}$ and $K_{1}$ are the reproducing kernels for the subspaces $\mathcal{H}_0$ and $\mathcal{H}_1$, respectively. Let $\{\xi_1, \ldots, \xi_{m}\}$ be a set of orthonormal basis functions for the finite-dimensional null space $\mathcal{H}_0$, where $m=\operatorname{dim}(\mathcal{H}_0)$. Then, the following theorem provides the closed-form representation of $\widehat{\beta}$.

\begin{theorem}
	\label{theo:beta_hat}
	The penalized maximum likelihood estimator for $\beta$ is 
	\begin{align*}
		\widehat{\beta}(s) = \sum_{j=1}^{m} d_j \xi_j(s) + \sum_{i=1}^n c_i \int_{\mathcal{S}} Z_i(w) K_{1}(w, s)\mathrm{d}w,
	\end{align*}
	where $d_j$ ($j=1, \ldots, m$) and $c_i$ ($i=1, \ldots, n$) are unknown constant coefficients to be estimated.
\end{theorem}

Theorem \ref{theo:beta_hat} enables computationally tractable estimation of $\beta$ by parameterizing it into a finite set of coefficients. Specifically, let $\boldsymbol{B}$ be an $m \times n$ matrix with its $(j,i)$-th element being $B_{ji} = \int_{\mathcal{S}}\xi_{j}(s)Z_{i}(s)\mathrm{d}s$ ($j=1,\ldots,m;i=1,\ldots,n$), and let $\boldsymbol{Q}$ be an $n \times n$ matrix with its $(i,j)$-th element being $Q_{ij} = \int_{\mathcal{S}}\int_{\mathcal{S}}Z_{i}(w)K_{1}(w,s)Z_{j}(s)\mathrm{d}w\mathrm{d}s$ ($i,j=1,\ldots,n$). Then, by Theorem \ref{theo:beta_hat},
\begin{align*}
	\int_{\mathcal{S}} \widehat{\beta}(s) Z_{i}(s) \mathrm{d}s = \boldsymbol{d}^{\top}\boldsymbol{B}_{i}+\boldsymbol{c}^{\top}\boldsymbol{Q}_{i} \quad \text { and } \quad J(\widehat{\beta}) = \boldsymbol{c}^{\top}\boldsymbol{Q}\boldsymbol{c},
\end{align*}
where $\boldsymbol{d}=(d_{1},\ldots,d_{m})^{\top}$, $\boldsymbol{c}=(c_{1},\ldots,c_{n})^{\top}$, and $\boldsymbol{B}_{i}$ and $\boldsymbol{Q}_{i}$ denote the $i$th columns of $\boldsymbol{B}$ and $\boldsymbol{Q}$, respectively. Thus, the estimation of infinite-dimensional $\beta$ is reduced to the estimation of the coefficient vectors $\boldsymbol{d}$ and $\boldsymbol{c}$.

The proof of Theorem \ref{theo:beta_hat} relies on an extension of the representer theorem of \citet{wahba1990spline} to the loglikelihood of interval-censored data. Analogous strategies for estimating the coefficient function have been adopted in functional linear models \citep{yuan2010, sun2018optimal} and functional Cox models with right-censored data \citep{qu2016optimal, hao2021semiparametric}. 

To make the aforementioned abstract framework concrete, we provide the following specific example.

\begin{example}
	\label{exm:sobolev}
	Assume without loss of generality that $\mathcal{S}=[0,1]$. A canonical choice of $\mathcal{H}$ is the Sobolev space of order $m$, defined as
	\begin{equation*}
		\mathcal{W}_{2, m} = \{f:[0,1] \rightarrow \mathbb{R} \mid f^{(0)}, f^{(1)}, \ldots, f^{(m-1)} \text{ are absolutely continuous, and\ }f^{(m)} \in L_2\},
	\end{equation*}
	where $f^{(j)}$ denotes the $j$th order derivative of $f$. According to Chapter 2.3 of \citet{gu2013smoothing},
	it is a reproducing kernel Hilbert space when endowed with the norm
	\begin{equation*}
		\|f\|_{\mathcal{W}_{2, m}} = \left[\sum_{j=1}^{m} \{f^{(j-1)}(0)\}^{2} + \int_{0}^{1}\{f^{(m)}(s)\}^2 \mathrm{d}s\right]^{1/2}.
	\end{equation*}
	Under this setup, the penalty function is naturally chosen as the squared $L_2$-norm of the $m$th derivative, namely $J(f)=\int_{0}^{1}\{f^{(m)}(s)\}^2 \mathrm{d}s$. Consequently, the null space $\mathcal{H}_0$ consists of polynomials of degree up to $m-1$ with the orthonormal basis functions $\xi_{j}(s)=s^{j-1}/(j-1)!$ ($j=1,\ldots,m$). Write $G_{m}(s, u)=(s-u)_{+}^{m-1} /(m-1)!$, where $(\cdot)_{+}=\max(0,\cdot)$. Then, the reproducing kernels for $\mathcal{H}_0$ and $\mathcal{H}_1$ are in forms of $K_{0}(w, s)=\sum_{j=1}^m \xi_{j}(w) \xi_{j}(s)$ and $K_{1}(w, s)=\int_{0}^{1} G_m(w, u) G_m(s, u) \mathrm{d}u$, respectively.
	Thus, the $(j,i)$-th element of $\boldsymbol{B}$ and the $(i,j)$-th element of $\boldsymbol{Q}$ are 
	\begin{equation*}
		B_{ji} = \int_{0}^{1} \frac{s^{j-1}}{(j-1)!} Z_i(s) \mathrm{d}s \quad \text { and } \quad Q_{ij} = \int_{0}^{1} \widetilde{Z}_{i}(u) \widetilde{Z}_{j}(u) \mathrm{d}u,
	\end{equation*}
	respectively, where $\widetilde{Z}_{i}(u) = \int_{0}^{1} Z_i(w) G_m(w, u) \mathrm{d}w$ serves as a transformed covariate.
\end{example}

\subsection{The EM Algorithm}
Define $L=\max\{U_{k}:U_{k}<T\}$ and $R=\min\{U_{k}:U_{k}\geq T\}$, and let $0=t_0<t_1<\cdots<t_q$ denote the set consisting of 0 and the unique values of $L_i>0$ and $R_i<\infty$ ($i=1, \ldots, n$).
We adopt the nonparametric maximum likelihood estimation approach, under which $\Lambda$ is regarded a step function with nonnegative jumps $\lambda_k$ at $t_k$ ($k=1, \ldots, q$). We define the concatenated parameter vector $\boldsymbol{\zeta} = (\boldsymbol{\alpha}^{\top}, \boldsymbol{d}^{\top}, \boldsymbol{c}^{\top})^{\top}$. Then, we maximize
\begin{align}
	\label{loglike}
	&n^{-1}\sum_{i=1}^{n}\log\left[\exp \left\{-\sum_{t_{k}\leq L_{i}} \lambda_{k}\exp(\boldsymbol{\zeta}^{\top}\widetilde{\boldsymbol{X}}_{ik}) \right\}-\exp \left\{-\sum_{t_{k}\leq R_{i}} \lambda_{k}\exp(\boldsymbol{\zeta}^{\top}\widetilde{\boldsymbol{X}}_{ik})\right\}\right]-\gamma_{n}\boldsymbol{c}^{\top}\boldsymbol{Q}\boldsymbol{c},
\end{align}
where $\widetilde{\boldsymbol{X}}_{ik}=\{\boldsymbol{X}_{i}(t_{k})^{\top}, \boldsymbol{B}_{i}^{\top}, \boldsymbol{Q}_{i}^{\top}\}^{\top}$. Direct maximization of the penalized loglikelihood in \eqref{loglike} is difficult, so we extend the EM algorithm of \citet{Zeng2016}. Specifically, we introduce mutually independent Poisson random variables $P_{ik}$ with means $\lambda_k \exp(\boldsymbol{\zeta}^{\top}\widetilde{\boldsymbol{X}}_{ik})$ $(i=1, \ldots, n; k=1, \ldots, q)$. In addition, let $N_{i1}=\sum_{t_{k} \leq L_{i}} P_{ik}$ and $N_{i2}=I(R_{i}<\infty) \sum_{L_{i}<t_{k} \leq R_{i}} P_{ik}$. The penalized loglikelihood for the observed data $\{N_{i1}=0, N_{i2}>0: i=1, \ldots, n\}$ is the same as \eqref{loglike}. Thus, we devise an EM algorithm by treating $P_{ik}$ as latent variables, where the M-step reduces to maximizing a penalized weighted sum of Poisson loglikelihood functions and admits closed-form solutions for $\lambda_{k}$ ($k=1,\ldots,q$).

In the E-step, we calculate the posterior mean of $P_{ik}$ given the observed data:
\begin{align*}
	\widehat{E}(P_{ik})=\frac{I(L_{i}<t_{k}\leq R_{i}<\infty)\lambda_{k}\exp(\boldsymbol{\zeta}^{\top}\widetilde{\boldsymbol{X}}_{ik})}{1-\exp\{\sum_{L_{i}<t_{k^{\prime}}\leq R_{i}}\lambda_{k^{\prime}}\exp(\boldsymbol{\zeta}^{\top}\widetilde{\boldsymbol{X}}_{ik^{\prime}})\}}.
\end{align*}
In the M-step, we maximize the following function via the one-step Newton–Raphson method to update $\boldsymbol{\zeta}$:
\begin{align}
	\label{EM_likelihood}
	&n^{-1} \sum_{i=1}^n \sum_{k=1}^q I(R_i^* \geq t_k)\widehat{E}(P_{ik})\Bigg[\boldsymbol{\zeta}^{\top}\widetilde{\boldsymbol{X}}_{ik}-\log \Bigg\{\sum_{i^{\prime}=1}^n I(R_{i^{\prime}}^* \geq t_k) \exp(\boldsymbol{\zeta}^{\top}\widetilde{\boldsymbol{X}}_{i^{\prime}k})\Bigg\}\Bigg]-\gamma_{n} \boldsymbol{c}^{\top}\boldsymbol{Q}\boldsymbol{c},
\end{align}
where $R_i^*=R_i I(R_i<\infty)+L_i I(R_i=\infty)$. We then update
\begin{align*}
	\lambda_{k}=\frac{\sum_{i=1}^{n} I(R_{i}^* \geq t_{k}) \widehat{E}(P_{ik})}{\sum_{i=1}^n I(R_{i}^* \geq t_{k})\exp(\boldsymbol{\zeta}^{\top}\widetilde{\boldsymbol{X}}_{ik})}
\end{align*}
for $k=1,\ldots,q$. Starting with $\boldsymbol{\zeta}=0$ and $\lambda_{k}=1 / q$, we iterate between the E-step and the M-step until the maximum absolute difference between the parameter estimates of two successive iterations is less than a certain threshold to obtain the estimators $\widehat{\boldsymbol{\zeta}}$ and $\widehat{\Lambda}$. 

This EM algorithm is computationally appealing because the parameters $\lambda_{k}$ are calculated explicitly in the M-step, circumventing the inversion of any high-dimensional matrices concerning these parameters. By Theorem \ref{theo:beta_hat}, the full parameter estimator $\widehat{\boldsymbol{\theta}}$ is uniquely determined by $\widehat{\boldsymbol{\zeta}}$ and $\widehat{\Lambda}$. With a slight abuse of notation, we write $l(\boldsymbol{\zeta},\Lambda)$ and $l(\boldsymbol{\theta})$ interchangeably in what follows.

\subsection{Tuning Parameter Selection}
The tuning parameter $\gamma_n$ is selected through an approximate leave-one-out cross-validation score. Let $\widehat{\boldsymbol{\zeta}}^{(-i)}$ and $\widehat{\Lambda}^{(-i)}$ denote the penalized estimators obtained by the EM algorithm with the $i$th subject's data excluded. The exact leave-one-out cross-validation score based on the observed loglikelihood is $-n^{-1} \sum_{i=1}^n l_i(\widehat{\boldsymbol{\zeta}}^{(-i)},\widehat{\Lambda}^{(-i)})$, where $l_{i}$ denotes the observed loglikelihood function for the $i$th subject. 

Because recalculating $\widehat{\boldsymbol{\zeta}}^{(-i)}$ and $\widehat{\Lambda}^{(-i)}$ for each subject is computationally prohibitive, we analytically approximate these leave-one-out estimators using the full-data estimators. Specifically, we consider the contribution of the $i$th subject to the unpenalized working loglikelihood in \eqref{EM_likelihood}, defined as
\begin{equation*}
	\widetilde{l}_i(\boldsymbol{\zeta}) = n^{-1}\sum_{k=1}^q I(R_i^* \geq t_k)\widehat{E}(P_{ik})\left[\boldsymbol{\zeta}^{\top}\widetilde{\boldsymbol{X}}_{ik}-\log \left\{\sum_{i^{\prime}=1}^n I(R_{i^{\prime}}^* \geq t_k) \exp(\boldsymbol{\zeta}^{\top}\widetilde{\boldsymbol{X}}_{i^{\prime}k})\right\}\right],
\end{equation*}
where $\widehat{E}(P_{ik})$ denotes the posterior mean of $P_{ik}$ evaluated at the converged full-data estimators $(\widehat{\boldsymbol{\zeta}},\widehat{\Lambda})$. Let $\widetilde{\boldsymbol{Q}} = \mathrm{diag}(\boldsymbol{0}_{p \times p}, \boldsymbol{0}_{m \times m}, \boldsymbol{Q})$ be the block-diagonal penalty matrix, where $0$ with a subscript denotes a zero matrix of the corresponding dimensions. Define $\widetilde{\boldsymbol{I}}_0 = -\sum_{i=1}^n \nabla_{\boldsymbol{\zeta}}^2 \widetilde{l}_i(\widehat{\boldsymbol{\zeta}})$ and $\widetilde{\boldsymbol{I}}_{\gamma_n} = \widetilde{\boldsymbol{I}}_0 + 2\gamma_n \widetilde{\boldsymbol{Q}}$ as the unpenalized and penalized working information matrices, respectively. 

By applying a first-order Taylor series expansion, we can obtain the approximation $\widehat{\boldsymbol{\zeta}}^{(-i)} \approx \widehat{\boldsymbol{\zeta}} - \widetilde{\boldsymbol{I}}_{\gamma_n}^{-1} \nabla_{\boldsymbol{\zeta}} \widetilde{l}_i(\widehat{\boldsymbol{\zeta}})$. The leave-one-out baseline hazard jumps are then approximated as
\begin{equation*}
	\widehat{\lambda}_{k}^{(-i)} \approx \widehat{\lambda}_{k} \left\{1 - \frac{\sum_{i^{\prime}=1}^n \widehat{w}_{i^{\prime}k} (\widehat{\boldsymbol{\zeta}}^{(-i)} - \widehat{\boldsymbol{\zeta}})^{\top}\widetilde{\boldsymbol{X}}_{i^{\prime}k}}{\sum_{i^{\prime}=1}^n \widehat{w}_{i^{\prime}k}}\right\} - \frac{I(R_{i}^* \geq t_{k}) \widehat{E}(P_{ik}) - \widehat{\lambda}_{k} \widehat{w}_{ik}}{\sum_{i^{\prime}=1}^n \widehat{w}_{i^{\prime}k}},
\end{equation*}
where $\widehat{w}_{ik} = I(R_{i}^* \geq t_{k})\exp(\widehat{\boldsymbol{\zeta}}^{\top}\widetilde{\boldsymbol{X}}_{ik})$. We approximate $\widehat{\Lambda}^{(-i)}$ by the step function constructed from the approximated $\widehat{\lambda}_k^{(-i)}$. Substituting these approximations into the exact cross-validation score yields the proposed approximate leave-one-out cross-validation score. We relegate the detailed derivation of these approximations to Section S.1 of the Supplementary Material.



\section{Theoretical Results}
\label{sec:theory}
\subsection{Score and Information Operators}
\label{sec:score-information}
Before establishing the asymptotic properties of $\widehat{\boldsymbol{\theta}}$, we derive its score and information operators, which play a key role in the consistency and weak convergence results. We consider the one-dimensional submodel $\epsilon \mapsto \boldsymbol{\theta}_{\epsilon,\boldsymbol{h}} \equiv \boldsymbol{\theta}+\epsilon\{\boldsymbol{h}_{1}, h_{2},\int_0^{(\cdot)} h_{3}(t) \mathrm{d}\Lambda(t)\}$, where $\boldsymbol{h}=(\boldsymbol{h}_{1},h_{2},h_{3})\in H\equiv\{\boldsymbol{f}=(\boldsymbol{f}_{1},f_{2},f_{3}):\boldsymbol{f}_{1}\in \mathbb{R}^{p}, f_{2}\in\mathcal{H}, f_{3}\in L_{2}[0,\tau]\}$. Then, the score operator for $\boldsymbol{\theta}$ is
\begin{align*}
	S_{\boldsymbol{\theta}}(\boldsymbol{h})\equiv \left.\frac{\partial}{\partial \epsilon} l(\boldsymbol{\theta}_{\epsilon,\boldsymbol{h}})\right|_{\epsilon=0}=\sum_{k=0}^M \Delta_k \int_{0}^{\tau} \{\boldsymbol{h}_{1}^{\top}\boldsymbol{X}(t)+\eta_{Z}(h_{2})+h_{3}(t)\}\Psi_{k}(t\mid \boldsymbol{\theta},\boldsymbol{X},Z)\mathrm{d} \Lambda(t),
\end{align*}
where $\eta_{Z}(h_{2})=\int_{\mathcal{S}}h_{2}(s)Z(s)\mathrm{d}s$,
\begin{align*}
	\begin{split}
		\Psi_{k}(t\mid\boldsymbol{\theta},\boldsymbol{X},Z)&=\exp\{\boldsymbol{\alpha}^{\top} \boldsymbol{X}(t)+\eta_{Z}(\beta)\}\\
		&\quad\quad\quad\quad\times\frac{S(U_{k+1}\mid\boldsymbol{\theta},\boldsymbol{X},Z) I(U_{k+1} \geq t)-S(U_{k}\mid\boldsymbol{\theta},\boldsymbol{X},Z) I(U_{k} \geq t)}{S(U_{k}\mid\boldsymbol{\theta},\boldsymbol{X},Z)-S(U_{k+1}\mid\boldsymbol{\theta},\boldsymbol{X},Z)}
	\end{split}
\end{align*}
for $k=0,\ldots,M$, and $S(t\mid \boldsymbol{\theta},\boldsymbol{X},Z)=\exp[-\int_{0}^{t}\exp\{\boldsymbol{\alpha}^{\top}\boldsymbol{X}(u)+\eta_{Z}(\beta)\}\mathrm{d}\Lambda(u)]$ is the survival function under model~\eqref{Cox}. Denote the expectation of the score operator by $S_{0\boldsymbol{\theta}}(\boldsymbol{h})\equiv\mathbb{P}\{S_{\boldsymbol{\theta}}(\boldsymbol{h})\}$, where $\mathbb{P}$ denotes the true probability measure. 

We equip $H$ with an inner product 
$\langle \boldsymbol{g},\boldsymbol{h} \rangle_{\Lambda} \equiv \boldsymbol{g}_{1}^{\top}\boldsymbol{h}_{1}+\int_{\mathcal{S}}g_{2}(s)h_{2}(s)\mathrm{d}s+\int_{0}^{\tau}g_{3}(t)h_{3}(t)\mathrm{d}\Lambda(t),$
so that it becomes a Hilbert space when $\Lambda$ is an increasing function. Differentiating $S_{0\boldsymbol{\theta}}(\boldsymbol{h})$ along the submodel $\epsilon \mapsto \boldsymbol{\theta}_{\epsilon,\boldsymbol{g}}=\boldsymbol{\theta}+\epsilon\{\boldsymbol{g}_{1},g_{2}, \int_0^{(\cdot)} g_{3}(t) \mathrm{d}\Lambda(t)\}$, we obtain
\begin{align} \label{eq:fisher-operator}
	I_{0\boldsymbol{\theta}}(\boldsymbol{g},\boldsymbol{h})\equiv-\left.\frac{\partial}{\partial \epsilon}S_{0\boldsymbol{\theta}}(\boldsymbol{\theta}_{\epsilon,\boldsymbol{g}})(\boldsymbol{h})\right|_{\epsilon=0}=\langle \boldsymbol{g}, \Gamma_{\boldsymbol{\theta}}(\boldsymbol{h})\rangle_{\Lambda},
\end{align}
where the linear operator $\Gamma_{\boldsymbol{\theta}}: H \rightarrow H$ is defined as the expectation of its empirical counterpart, namely $\Gamma_{\boldsymbol{\theta}}(\boldsymbol{h}) = \mathbb{P}\widehat{\Gamma}_{\boldsymbol{\theta}}(\boldsymbol{h})$. Here, $\widehat{\Gamma}_{\boldsymbol{\theta}} = \sum_{k=0}^{M} \Delta_{k} \boldsymbol{V}_{k}(\boldsymbol{\theta})^{\otimes 2}$, where $\boldsymbol{a}^{\otimes 2} = \boldsymbol{a}\boldsymbol{a}^{\top}$ for any vector $\boldsymbol{a}$,
\begin{equation*}
	\boldsymbol{V}_{k}(\boldsymbol{\theta}) = 
	\begin{bmatrix} 
		\int_0^{\tau} \boldsymbol{X}(u) D_{k}(u \mid \boldsymbol{\theta}, \boldsymbol{X}, Z) \mathrm{d}\Lambda(u) \\[1ex] 
		\int_0^{\tau} Z(\cdot) D_{k}(u \mid \boldsymbol{\theta}, \boldsymbol{X}, Z) \mathrm{d}\Lambda(u) \\[1ex] 
		D_{k}(\cdot \mid \boldsymbol{\theta}, \boldsymbol{X}, Z) 
	\end{bmatrix},
\end{equation*}
and
\begin{equation*}
	D_{k}(t\mid\boldsymbol{\theta},\boldsymbol{X},Z) = I(U_{k}<t\leq U_{k+1})\exp\{\boldsymbol{\alpha}^{\top} \boldsymbol{X}(t)+\eta_{Z}(\beta)\}\frac{S^{1/2}(U_{k}\mid\boldsymbol{\theta},\boldsymbol{X},Z)S^{1/2}(U_{k+1}\mid\boldsymbol{\theta},\boldsymbol{X},Z)}{S(U_{k}\mid\boldsymbol{\theta},\boldsymbol{X},Z)-S(U_{k+1}\mid\boldsymbol{\theta},\boldsymbol{X},Z)}.
\end{equation*}
The action of $\widehat{\Gamma}_{\boldsymbol{\theta}}$ on $\boldsymbol{h}\in H$ is defined as $\widehat{\Gamma}_{\boldsymbol{\theta}}(\boldsymbol{h})=\sum_{k=0}^{M} \Delta_{k} \boldsymbol{V}_{k}(\boldsymbol{\theta})\langle \boldsymbol{V}_{k}(\boldsymbol{\theta}),\boldsymbol{h} \rangle_{\Lambda}$. 
In addition, $\Gamma_{\boldsymbol{\theta}}(\boldsymbol{h})$ and $\widehat{\Gamma}_{\boldsymbol{\theta}}(\boldsymbol{h})$ can be partitioned as
\begin{equation*}
	\Gamma_{\boldsymbol{\theta}}(\boldsymbol{h}) = \mathbb{P} \widehat{\Gamma}_{\boldsymbol{\theta}}(\boldsymbol{h}) =
	\mathbb{P}\begin{bmatrix}
		\widehat{\Gamma}_{\boldsymbol{\theta}}^{11} & \widehat{\Gamma}_{\boldsymbol{\theta}}^{12} & \widehat{\Gamma}_{\boldsymbol{\theta}}^{13}\\
		\widehat{\Gamma}_{\boldsymbol{\theta}}^{21} & \widehat{\Gamma}_{\boldsymbol{\theta}}^{22} & \widehat{\Gamma}_{\boldsymbol{\theta}}^{23}\\
		\widehat{\Gamma}_{\boldsymbol{\theta}}^{31} & \widehat{\Gamma}_{\boldsymbol{\theta}}^{32} & \widehat{\Gamma}_{\boldsymbol{\theta}}^{33}
	\end{bmatrix}
	\begin{bmatrix}
		\boldsymbol{h}_{1} \\
		h_2 \\
		h_3
	\end{bmatrix}
	= 
	\begin{bmatrix}
		\Gamma_{\boldsymbol{\theta}}^{11} & \Gamma_{\boldsymbol{\theta}}^{12} & \Gamma_{\boldsymbol{\theta}}^{13}\\
		\Gamma_{\boldsymbol{\theta}}^{21} & \Gamma_{\boldsymbol{\theta}}^{22} & \Gamma_{\boldsymbol{\theta}}^{23}\\
		\Gamma_{\boldsymbol{\theta}}^{31} & \Gamma_{\boldsymbol{\theta}}^{32} & \Gamma_{\boldsymbol{\theta}}^{33}
	\end{bmatrix}
	\begin{bmatrix}
		\boldsymbol{h}_{1} \\
		h_2 \\
		h_3
	\end{bmatrix},
\end{equation*}
where each block operator $\Gamma_{\boldsymbol{\theta}}^{ij} = \mathbb{P}(\widehat{\Gamma}_{\boldsymbol{\theta}}^{ij})$ corresponds to the expectation of the $(i,j)$-th block of $\sum_{k=0}^{M} \Delta_{k} \boldsymbol{V}_{k}(\boldsymbol{\theta})^{\otimes 2}$ for $i,j\in\{1,2,3\}$.

\subsection{Regularity Conditions}
We establish the asymptotic properties of $\widehat{\boldsymbol{\theta}}$ under the following regularity conditions.
\begin{condition} \label{cond:comp-para}
	The true value of $\boldsymbol{\alpha}$, denoted by $\boldsymbol{\alpha}_0$, lies in the interior of a known compact set in $\mathbb{R}^{p}$. The true value of $\beta$, denoted by $\beta_0$, lies in a bounded subset of $\mathcal{H}$. Finally, the true value of $\Lambda$, denoted by $\Lambda_0$, is  continuously differentiable with positive derivatives on $[0, \tau]$. 
\end{condition}

\begin{condition} \label{cond:bounded}
	The covariate process $\boldsymbol{X}(t)$ is uniformly bounded with uniformly bounded total variation over $[0, \tau]$, and the $L_2$-norm of the functional covariate $Z(s)$ over $\mathcal{S}$ is uniformly bounded almost surely, where $\mathcal{S}$ is a compact set.
\end{condition}

\begin{condition} \label{cond:identification}
	If there exist a constant vector $\boldsymbol{f}_{1}$ and deterministic functions $f_{2}$ and $f_{3}$ such that  $\boldsymbol{f}_{1}^{\top} \boldsymbol{X}(t) + \int_{\mathcal{S}} f_{2}(s) Z(s) \mathrm{d}s + f_{3}(t) = 0$ for all $t \in [0, \tau]$ with probability 1, then $\boldsymbol{f}_{1} = 0$, $f_{2}(s) = 0$ for $s\in\mathcal{S}$, and $f_{3}(t) = 0$ for $t \in [0, \tau]$. 
\end{condition} 

\begin{condition} \label{cond:U&M}
	The number of examination times $M$ is positive with $E(M)<\infty$. The probability $\mathrm{pr}(U_{M}=\tau \mid M, \boldsymbol{X}, Z)$ is greater than some positive constant. In addition, $\mathrm{pr}\{\min _{0 \leq k \leq M}(U_{k+1}-U_{k}) \geq c_{0} \mid M, \boldsymbol{X}, Z\}=1$ for some positive constant $c_{0}$. Finally, the densities of $(U_{k}, U_{k+1})$ conditional on $(M, \boldsymbol{X}, Z)$, denoted by $g_{k}(u, v\mid M, \boldsymbol{X}, Z)$ ($k=0, \ldots, M$), have continuous second-order partial derivatives with respect to $u$ and $v$ when $v-u \geq c_{0}$ and $u, v \in[0, \tau]$, and they are continuously differentiable functionals with respect to $\boldsymbol{X}$ and $Z$.
\end{condition}

\begin{condition} \label{cond:kernel}
	The reproducing kernel $K(\cdot,\cdot)$ is continuous and uniformly bounded on $\mathcal{S} \times \mathcal{S}$, such that $\sup_{s \in \mathcal{S}} K(s,s) < \infty$. In addition, the eigenvalues associated with the spectral decomposition of $K$ satisfy $\kappa_{j} \lesssim j^{-2r}$ for some constant $r \geq 1$, where $x_j \lesssim y_j$ indicates that there exists a universal constant $C_{0} > 0$ independent of $j$ such that $x_{j} \leq C_{0} y_{j}$ for all $j \in \mathbb{N}^{+}$.
\end{condition}

Condition \ref{cond:comp-para} imposes basic compactness and smoothness on the parameter spaces. Condition \ref{cond:bounded} imposes boundedness restrictions on the covariates and allows $\boldsymbol{X}$ and $Z$ to have discontinuous trajectories. This condition is used to prove the Donsker property of the involved function classes. Condition \ref{cond:identification} is necessary to establish the identifiability of the model parameters. This condition naturally holds provided that the scalar covariates, the functional covariate, and time are not perfectly linearly dependent and exhibit sufficient variations. Condition \ref{cond:U&M} requires the separation of two adjacent examination times by at least a constant $c_{0}$, along with the smoothness of the joint distribution of the examination times. Conditions \ref{cond:comp-para}--\ref{cond:U&M} are analogous to Conditions 1--4 of \citet{Zeng2016}, apart from the additional regularities imposed on the coefficient function and the functional covariate, which are standard in the functional regression literature \citep[e.g.,][]{ma2005penalized,qu2016optimal,hao2021semiparametric}.

Finally, the first part of Condition \ref{cond:kernel} guarantees a bounded $L_\infty$-norm for bounded functions in $\mathcal{H}$, and the second part requires the eigenvalues to decay at a polynomial rate. This decay rate inherently quantifies the topological complexity of the parameter space and bounds its bracketing entropy, which guarantees the convergence of Dudley’s entropy integral and establishes the Donsker property required for the asymptotic theory. Condition \ref{cond:kernel} accommodates a broad class of kernels, including the Sobolev kernel in Example \ref{exm:sobolev} \citep{micchelli1981design} and the Gaussian radial basis function kernel \citep{rasmussen2006gaussian}.

\subsection{Asymptotic Properties}
We state the strong consistency and the overall convergence rate of $\widehat{\boldsymbol{\theta}}$ in Theorems \ref{thm:consistency} and \ref{thm:rate}, respectively. 
\begin{theorem} \label{thm:consistency}
	Suppose that Conditions \ref{cond:comp-para}--\ref{cond:kernel} hold. If $\gamma_n =o(1)$, then
	\begin{equation*}
		\|\widehat{\boldsymbol{\alpha}} - \boldsymbol{\alpha}_0\|_2 + \sup_{s \in \mathcal{S}} |\widehat{\beta}(s) - \beta_0(s)| + \sup_{t \in [0, \tau]} |\widehat{\Lambda}(t) - \Lambda_0(t)| \rightarrow 0
	\end{equation*}
	almost surely, where $\|\cdot\|_2$ is the Euclidean norm.
\end{theorem}
\begin{theorem} \label{thm:rate}
	Suppose that Conditions \ref{cond:comp-para}--\ref{cond:kernel} hold. If $\gamma_{n} = O(n^{-2/3})$,  
	then $\mathcal{D}_{n} =  O_p(n^{-1/3}) $, where
	\begin{align*}
		\mathcal{D}_{n} = E\left(\sum_{k = 1}^M \left[\int_0^{U_k} \exp\{\widehat{\boldsymbol{\alpha}}^{\top} \boldsymbol{X}(t) + \eta_{Z}(\widehat{\beta})\} \mathrm{d} \widehat{\Lambda}(t) - \int_0^{U_k} \exp\{\boldsymbol{\alpha}_{0}^{\top} \boldsymbol{X}(t) + \eta_{Z}(\beta_{0})\} \mathrm{d} {\Lambda}_0(t) \right]^2\right)^{1/2}.
	\end{align*}
\end{theorem}

Theorem \ref{thm:rate} establishes the convergence rate of $\widehat{\boldsymbol{\theta}}$ in terms of a predictive risk metric. Specifically, it bounds the expected $L_{2}$-distance between the estimated and the true subject-specific cumulative hazard functions, evaluated over a new sequence of examination times. The overall convergence rate of $\widehat{\boldsymbol{\theta}}$ is dominated by $\widehat{\Lambda}$, whose $n^{1/3}$-convergence rate matches the minimax rate of estimation over the class of uniformly bounded functions with uniformly bounded total variation \citep{yang1999minimax}. Because this nonparametric rate is slower than the parametric $n^{1/2}$ rate and precludes standard techniques for deriving the joint asymptotic distribution of $(\widehat{\boldsymbol{\alpha}},\widehat{\beta})$ directly, we instead evaluate the estimators through smooth linear functionals. This approach allows us to establish the Hoffmann--J{\o}rgensen weak convergence \citep{van1996weak} of the resulting stochastic process.

Let $\mathcal{V}_{\boldsymbol{\alpha}}$ denote the unit ball in $\mathbb{R}^p$, and $\mathcal{V}_{\beta}$ denote the unit ball in the functional space $\mathcal{H}$ consisting of functions with an intrinsic norm bounded by 1. Write $\mathcal{V} = \mathcal{V}_{\boldsymbol{\alpha}} \times \mathcal{V}_{\beta}$. For any specific direction $\boldsymbol{f}=(\boldsymbol{f}_{1}, f_{2}) \in \mathcal{V}$, we define the joint linear evaluation functional $\Phi_n(\boldsymbol{f})$ as
\begin{equation*}
	\Phi_n(\boldsymbol{f}) = \boldsymbol{f}_{1}^{\top}(\widehat{\boldsymbol{\alpha}}-\boldsymbol{\alpha}_{0}) + \int_{\mathcal{S}} f_{2}(s)\{\widehat{\beta}(s)-\beta_{0}(s)\}\mathrm{d}s.
\end{equation*}
We treat the stochastic process $\Phi_n$ as a random element in the metric space $l^{\infty}(\mathcal{V})$, which consists of all uniformly bounded real-valued functionals defined on $\mathcal{V}$. Then, the weak convergence of $\Phi_n$ is established in the following theorem.
\begin{theorem} \label{thm:weakconv}
	Under Conditions \ref{cond:comp-para}--\ref{cond:kernel} and $\gamma_{n} = O(n^{-2/3})$, the stochastic process $n^{1/2} \Phi_n$ converges weakly to a zero-mean Gaussian process in the metric space $l^{\infty}(\mathcal{V})$. In addition, for any specific direction $\boldsymbol{f}\in\mathcal{V}$, the asymptotic variance of $n^{1/2} \Phi_n(\boldsymbol{f})$ attains the semiparametric efficiency bound.
\end{theorem}

Theorem \ref{thm:weakconv} provides a unified framework for analysing the asymptotic properties of $(\widehat{\boldsymbol{\alpha}},\widehat{\beta})$. The significance of this theorem lies in its ability to simultaneously establish the $n^{1/2}$-consistency and semiparametric efficiency of the evaluated functionals, despite the slow $n^{1/3}$-convergence rate of $\widehat{\Lambda}$. Because this theorem is established in the general metric space $l^{\infty}(\mathcal{V})$, it serves as a master theorem from which various asymptotic distributions of interest can be readily derived. By selecting specific test directions $\boldsymbol{f}\in \mathcal{V}$, one can project $\Phi_{n}$ onto relevant subspaces to obtain the marginal or joint asymptotic distributions of the estimators for certain parameters, including the finite-dimensional $\widehat{\boldsymbol{\alpha}}$ and smooth functionals of $\widehat{\beta}$. Specifically, we have the following results.

\begin{corollary} \label{thm:semip-eff}
	Under Conditions \ref{cond:comp-para}--\ref{cond:kernel} and $\gamma_{n} = O(n^{-2/3})$, $n^{1/2}(\widehat{\boldsymbol{\alpha}} - \boldsymbol{\alpha}_{0})$ converges in distribution to a zero-mean $p$-variate normal random vector whose covariance matrix $\boldsymbol{\Sigma}$ attains the semiparametric efficiency bound. Specifically, $\boldsymbol{\Sigma} = [\mathbb{P}\{\boldsymbol{S}_{\boldsymbol{\theta}_{0}}(\boldsymbol{I}_{p},-\boldsymbol{h}_{2}^{*},-\boldsymbol{h}_{3}^{*})^{\otimes 2}\}]^{-1}$, where $\boldsymbol{\theta}_{0}=(\boldsymbol{\alpha}_{0},\beta_{0},\Lambda_{0})$, and
	\begin{equation*} \label{eq:effi-score}
		\boldsymbol{S}_{\boldsymbol{\theta}_{0}}(\boldsymbol{I}_{p},-\boldsymbol{h}_{2}^{*},-\boldsymbol{h}_{3}^{*}) = \sum_{k=0}^M \Delta_k \int_{0}^{\tau} \{\boldsymbol{X}(t) - \eta_{Z}(\boldsymbol{h}_{2}^{*}) - \boldsymbol{h}_{3}^{*}(t)\} \Psi_{k}(t \mid \boldsymbol{\theta}_{0}, \boldsymbol{X}, Z) \mathrm{d}\Lambda_0(t)
	\end{equation*}
	is the efficient score function for $\boldsymbol{\alpha}$. Here, $(\boldsymbol{h}_{2}^{*}, \boldsymbol{h}_{3}^{*}) \in \mathcal{H}^p \times L_{2}^p[0, \tau]$ is the least favourable direction that solves the operator equations $\Gamma_{\boldsymbol{\theta}_{0}}^{21}(\boldsymbol{I}_{p}) = \Gamma_{\boldsymbol{\theta}_{0}}^{22}(\boldsymbol{h}_{2}^{*}) + \Gamma_{\boldsymbol{\theta}_{0}}^{23}(\boldsymbol{h}_{3}^{*})$ and $\Gamma_{\boldsymbol{\theta}_{0}}^{31}(\boldsymbol{I}_{p}) = \Gamma_{\boldsymbol{\theta}_{0}}^{32}(\boldsymbol{h}_{2}^{*}) + \Gamma_{\boldsymbol{\theta}_{0}}^{33}(\boldsymbol{h}_{3}^{*})$ with $\Gamma_{\boldsymbol{\theta}_{0}}$ defined in \eqref{eq:fisher-operator}, and $\boldsymbol{I}_{p}$ denotes the $p \times p$ identity matrix.
\end{corollary}

\begin{corollary} \label{thm:beta-functional-asymp}
	Suppose that Conditions \ref{cond:comp-para}--\ref{cond:kernel} hold and that $\gamma_{n} = O(n^{-2/3})$. For any function $b \in \mathcal{H}$, the linear functional $n^{1/2} \int_{\mathcal{S}} b(s)\{\widehat{\beta}(s) - \beta_0(s)\} \mathrm{d}s$ converges in distribution to a zero-mean normal random variable whose variance is $\sigma_{b}^{2} = \mathbb{P} \{S_{\boldsymbol{\theta}_{0}}(\boldsymbol{h}_{b}^{*})^2\}$ and attains the semiparametric efficiency bound. Here, $\boldsymbol{h}_{b}^{*} \in H$ is the unique solution to the operator equation $\Gamma_{\boldsymbol{\theta}_{0}}(\boldsymbol{h}_{b}^{*}) = (0, b, 0)^{\top}$.
\end{corollary}

Corollaries \ref{thm:semip-eff} and \ref{thm:beta-functional-asymp} establish the asymptotic normality of the proposed estimators and provide theoretical justification for performing statistical inference on both the regression parameter $\boldsymbol{\alpha}$ and smooth functionals of $\beta(\cdot)$. Corollary \ref{thm:semip-eff} confirms that $\widehat{\boldsymbol{\alpha}}$ is asymptotically efficient among all regular estimators for $\boldsymbol{\alpha}$. Corollary \ref{thm:beta-functional-asymp} presents an analytical tool that establishes a theoretically sound foundation for inference on the infinite-dimensional coefficient function $\beta$. Specifically, by selecting a finite set of smooth test functions $b(s)$ and estimating the covariance matrix of their corresponding evaluated functionals, one can leverage this result to construct a Wald-type test statistic for formally evaluating the global null hypothesis $H_0: \beta(s) = 0$ for all $s \in \mathcal{S}$.

The theoretical contribution of Corollary \ref{thm:beta-functional-asymp} extends beyond the existing literature, as a comparable asymptotic result has not yet been established even for the simpler functional Cox model with right-censored data. \citet{qu2016optimal} focused primarily on estimation and did not develop inference procedures for $\beta$, and the inference framework proposed by \citet{hao2021semiparametric} relied on stringent conditions such as simultaneous diagonalization of the weighted covariance operator and the penalty function. In contrast, our approach bypasses these restrictive conditions by exploiting the bounded invertibility of the information operator. Furthermore, we extend the Wald-type test implied by Corollary \ref{thm:beta-functional-asymp} by allowing the number of test functions to diverge with the sample size to facilitate the global test for $\beta$, as stated in Theorem \ref{thm:chiappro} below.

\begin{theorem} \label{thm:chiappro}
	Suppose that Conditions \ref{cond:comp-para}--\ref{cond:kernel} hold and that $\gamma_{n} = O(n^{-2/3})$. Let $\{b_{l}\}_{l=1}^{\infty} \subset \mathcal{H}$ be a sequence of linearly independent test functions. Write $\boldsymbol{v}_{n}=(v_{n 1}, \ldots, v_{n L_n})^{\top}$, where $v_{nl}= n^{1/2} \int_{\mathcal{S}} b_{l}(s)\{\widehat{\beta}(s) - \beta_0(s)\}\mathrm{d}s$, and $L_{n}$ is a positive integer that may depend on $n$. Then, provided that $L_n = o(n^{1/6})$ as $n\rightarrow\infty$,
	\begin{equation} \label{eq:chiappro}
		\lim_{n \rightarrow \infty} \sup_{x\geq 0}\left|\mathrm{pr}(\boldsymbol{v}_{n}^{\top} \boldsymbol{\Omega}_{n}^{-1} \boldsymbol{v}_{n} \leq x) - \mathrm{pr}(W_n \leq x) \right| = 0,
	\end{equation}
	where $\boldsymbol{\Omega}_n$ is the $L_n \times L_n$ matrix with its $(j, l)$-th element given by $\mathbb{P} \{S_{\boldsymbol{\theta}_{0}}(\boldsymbol{h}_{b_j}^{*})S_{\boldsymbol{\theta}_{0}}(\boldsymbol{h}_{b_l}^{*})\}$, $\boldsymbol{h}_{b_l}^* \in H$ is the unique solution to the operator equation $\Gamma_{\boldsymbol{\theta}_{0}}(\boldsymbol{h}_{b_{l}}^{*}) = (0, b_{l}, 0)^{\top}$, and $W_n$ denotes a random variable following a chi-squared distribution with $L_n$ degrees of freedom.
\end{theorem}

\section{Statistical Inference}
\label{sec:infer}

To conduct valid statistical inference for both $\boldsymbol{\alpha}$ and $\beta$, it is essential to estimate the associated covariance matrices, $\boldsymbol{\Sigma}$ and $\boldsymbol{\Omega}_{n}$. We achieve this using the profile likelihood method \citep{profile}. Denote the dimension of $\boldsymbol{\zeta}$ by $N_{\boldsymbol{\zeta}}$. We consider a general inference framework for a target parameter $\boldsymbol{\rho} \in \mathbb{R}^{N_{\boldsymbol{\rho}}}$ defined by a linear transformation $\boldsymbol{\rho} = \boldsymbol{A} \boldsymbol{\zeta}$, where $\boldsymbol{A}$ is a prespecified $N_{\boldsymbol{\rho}} \times N_{\boldsymbol{\zeta}}$ matrix of rank $N_{\boldsymbol{\rho}} $ ($N_{\boldsymbol{\rho}} < N_{\boldsymbol{\zeta}})$. The penalized profile loglikelihood function for $\boldsymbol{\rho}$ is defined as $\mathrm{pl}_n(\boldsymbol{\rho}) = \max_{\boldsymbol{\zeta}, \Lambda} [ \mathbb{P}_{n} \{l(\boldsymbol{\zeta}, \Lambda)\} - \gamma_{n} \boldsymbol{c}^{\top}\boldsymbol{Q}\boldsymbol{c}]$ subject to $\boldsymbol{A} \boldsymbol{\zeta} = \boldsymbol{\rho}$.

Evaluating $\mathrm{pl}_n(\boldsymbol{\rho})$ requires maximizing the penalized loglikelihood function under the constraint $\boldsymbol{A} \boldsymbol{\zeta} = \boldsymbol{\rho}$, which can be efficiently solved by tailoring the EM algorithm described in Section \ref{sec:est}. The E-step and the procedure for updating $\lambda_{k}$ in the M-step remain identical to those in the unconstrained setting. To update $\boldsymbol{\zeta}$ under the constraint, we embed a null-space projection into the Newton--Raphson step. Let $\boldsymbol{A}_{0}$ be an $N_{\boldsymbol{\zeta}} \times (N_{\boldsymbol{\zeta}} - N_{\boldsymbol{\rho}} )$ matrix whose columns form an orthonormal basis for the null space of $\boldsymbol{A}$, satisfying $\boldsymbol{A} \boldsymbol{A}_{0} = 0$. An initial feasible iterate $\boldsymbol{\zeta}^{(0)}$ satisfying $\boldsymbol{A} \boldsymbol{\zeta}^{(0)} = \boldsymbol{\rho}$ can be obtained by projecting the unconstrained estimate $\widehat{\boldsymbol{\zeta}}$ onto the constraint hyperplane, yielding $\boldsymbol{\zeta}^{(0)} = \widehat{\boldsymbol{\zeta}} + \boldsymbol{A}^{\top}(\boldsymbol{A}\boldsymbol{A}^{\top})^{-1}(\boldsymbol{\rho} - \boldsymbol{A}\widehat{\boldsymbol{\zeta}})$. Given a feasible iterate $\boldsymbol{\zeta}^{(s)}$ at the $s$th iteration, the constrained Newton--Raphson update restricts the search direction to the column space of $\boldsymbol{A}_{0}$. By projecting the standard unconstrained parameter update onto the null space of $\boldsymbol{A}$, we obtain the subsequent iterate $\boldsymbol{\zeta}^{(s+1)}$, which algebraically guarantees that $\boldsymbol{A} \boldsymbol{\zeta}^{(s+1)} = \boldsymbol{\rho}$ for all $s$. Then, the covariance matrix of $n^{1/2}(\widehat{\boldsymbol{\rho}} - \boldsymbol{\rho})$ is estimated by $\{-\nabla_{\boldsymbol{\rho}}^2 \mathrm{pl}_n(\widehat{\boldsymbol{\rho}})\}^{-1}$, where $\widehat{\boldsymbol{\rho}} = \boldsymbol{A}\widehat{\boldsymbol{\zeta}}$, $\widehat{\boldsymbol{\zeta}}$ is the estimator for $\boldsymbol{\zeta}$ obtained in Section \ref{sec:est}, and $\nabla_{\boldsymbol{\rho}}^2 \mathrm{pl}_n(\widehat{\boldsymbol{\rho}})$ is the Hessian matrix of $\mathrm{pl}_n(\boldsymbol{\rho})$ evaluated at $\widehat{\boldsymbol{\rho}}$. Because there is no analytical expression for $\nabla_{\boldsymbol{\rho}}^2 \mathrm{pl}_n(\widehat{\boldsymbol{\rho}})$, we approximate its $(i,j)$-th element by
\begin{align*}
	\frac{\mathrm{pl}_n(\widehat{\boldsymbol{\rho}}) - \mathrm{pl}_n(\widehat{\boldsymbol{\rho}} + h_n \boldsymbol{e}_i) - \mathrm{pl}_n(\widehat{\boldsymbol{\rho}} + h_n \boldsymbol{e}_j) + \mathrm{pl}_n(\widehat{\boldsymbol{\rho}} + h_n \boldsymbol{e}_i + h_n \boldsymbol{e}_j)}{h_n^2},
\end{align*}
where $\boldsymbol{e}_i$ is the $i$th canonical vector in $\mathbb{R}^{N_{\boldsymbol{\rho}}}$, and $h_n$ is a perturbation constant of order $n^{-1/2}$.

This unified formulation naturally yields the estimators for both $\boldsymbol{\Sigma}$ and $\boldsymbol{\Omega}_{n}$. To estimate $\boldsymbol{\Sigma}$, we set the target parameter to $\boldsymbol{\rho} = \boldsymbol{\alpha}$ with the constraint matrix $\boldsymbol{A} = [\boldsymbol{I}_{p}, \boldsymbol{0}_{p \times (m+n)}]$. To perform the global test for $\beta(\cdot)$ established in Theorem \ref{thm:chiappro}, the target parameter becomes the evaluated functional vector $\boldsymbol{\rho} = (\rho_1, \ldots, \rho_{L_n})^{\top}$, where $\rho_l = \int_{\mathcal{S}} b_l(s) \beta(s) \mathrm{d}s$. Based on the closed-form representation in Theorem \ref{theo:beta_hat}, the corresponding constraint matrix $\boldsymbol{A}$ is constructed with the $l$th row being $(\boldsymbol{0}_{1 \times p}, \boldsymbol{u}_{ld}^{\top}, \boldsymbol{u}_{lc}^{\top})$, where $\boldsymbol{u}_{ld}$ is an $m$-vector with the $j$th element being $\int_{\mathcal{S}} \xi_{j}(s) b_l(s) \mathrm{d}s$, and $\boldsymbol{u}_{lc}$ is an $n$-vector with the $i$th element being $\int_{\mathcal{S}}\int_{\mathcal{S}} Z_{i}(w)K_{1}(w,s) b_l(s) \mathrm{d}w\mathrm{d}s$. Denote the resulting estimator for $\boldsymbol{\Omega}_n$ by $\widehat{\boldsymbol{\Omega}}_n$. The Wald-type statistic for testing $H_0: \beta(s) = 0$ for all $s \in \mathcal{S}$ is given by $n\widehat{\boldsymbol{\rho}}^{\top} \widehat{\boldsymbol{\Omega}}_n^{-1} \widehat{\boldsymbol{\rho}}$, where $\widehat{\boldsymbol{\rho}} = (\widehat{\rho}_1, \ldots, \widehat{\rho}_{L_n})^{\top}$ and $\widehat{\rho}_l = \int_{\mathcal{S}} b_l(s) \widehat{\beta}(s) \mathrm{d}s$. The corresponding $p$-value can be computed using the chi-squared reference distribution with $L_n$ degrees of freedom.

\section{Simulation Studies} \label{sec:simu}
We conducted simulation studies to evaluate the performance of the proposed methods. We simulated 1000 replicated datasets with sample sizes of 100, 200 or 400. For each subject, we generated two independent covariates, $X_{1}\sim \mathrm{Ber}(0.5)$ and $X_{2}\sim \mathrm{Unif}(0,1)$, and set the corresponding regression parameters $\alpha_{01}$ and $\alpha_{02}$ to 1 and $-$0.5, respectively. To generate the functional covariate $Z(s)$, we adopted a similar design to that of \citet{qu2016optimal}. Specifically, we set the domain to $\mathcal{S}=[0,1]$ and considered the functional principal components defined by $\psi_{1}(s)=1$ and $\psi_{j+1}(s)=\sqrt{2} \cos (j \pi s)$ for $j\geq 1$. We then generated $Z(s)=\sum_{j=1}^{50}(-1)^{j+1} j^{-v / 2}Y_{j}\psi_{j}(s)$, where the random coefficients $Y_{j}$ were independently drawn from $\mathrm{Unif}(-3,3)$, and $v \in \{1, 2, 3\}$ determines the decay rate of the eigenvalues of the covariance operator of $Z(\cdot)$. We set the corresponding coefficient function to $\beta_{0}(s)=\sum_{j=1}^{50}(-1)^{j} j^{-3 / 2} \psi_j(s)$. Then, we generated the event time $T$ based on the cumulative hazard function $\Lambda_{0}(t)\exp\{\alpha_{01}X_{1}+\alpha_{02}X_{2}+\eta_{Z}(\beta_{0})\}$, where $\Lambda_{0}(t)=0.25t^{2}$. We generated six potential examination times by generating six time points from $\mathrm{Unif}(0,5)$ and letting $U_{ik}$ be the $k$th ordered time point plus $\mathrm{Unif}\{0.1(k-1), 0.1k\}$. On average, there were around 27\% left-censored observations and 14\% right-censored ones across different $v$. Finally, we adopted the Sobolev space described in Example \ref{exm:sobolev} with $m=2$ and fit model \eqref{Cox} to each dataset with the convergence threshold in the EM algorithm set at $5\times10^{-3}$.

Table \ref{tab:alpha_sim} summarizes the estimation results for the regression parameters under different $v$. The proposed estimator for $\boldsymbol{\alpha}$ has small bias that decreases as $n$ increases. The empirical standard errors and the means of the estimated standard errors agree closely, indicating that the proposed variance estimator is highly accurate. The empirical coverage percentages of the corresponding confidence intervals are well-maintained around the nominal 95\% level. These favourable finite-sample properties remain robust across different decay rates $v$. 

\begin{table}[!ht]
	\centering
	\caption{Parameter estimation results for the regression parameters.}
	{%
		\scalebox{0.8}{
			\begin{threeparttable}
				\begin{tabular}{clrcccrcccrccc}
					\toprule
					& & \multicolumn{4}{c}{$v=1$}  & \multicolumn{4}{c}{$v=2$} & \multicolumn{4}{c}{$v=3$} \\
					\cmidrule(lr){3-6} \cmidrule(lr){7-10} \cmidrule(lr){11-14} 
					$n$ & Parameter & Bias & SE & SEE & CP & Bias & SE & SEE & CP & Bias & SE & SEE & CP \\
					\midrule
					100 & $\alpha_{01}=1$   
					& 0.193 & 0.360 & 0.375 & 95 & 0.168 & 0.359 & 0.361 & 95 & 0.168 & 0.346 & 0.354 & 94 \\
					& $\alpha_{02}=-0.5$ \vspace*{4pt}
					& $-$0.090 & 0.578 & 0.592 & 95 & $-$0.092 & 0.583 & 0.580 & 95 & $-$0.096 & 0.570 & 0.568 & 95\\
					
					200 & $\alpha_{01}=1$   
					& 0.093 & 0.228 & 0.231 & 94 & 0.096 & 0.230 & 0.225 & 94 & 0.102 & 0.236 & 0.225 & 94\\
					& $\alpha_{02}=-0.5$ \vspace*{4pt}
					& $-$0.040 & 0.362 & 0.373 & 95 & $-$0.056 & 0.359 & 0.369 & 95 & $-$0.052 & 0.365 & 0.363 & 95\\
					
					400 & $\alpha_{01}=1$   
					& 0.061 & 0.151 & 0.152 & 94 & 0.051 & 0.149 & 0.149 & 94 & 0.050 & 0.146 & 0.148 & 95\\
					& $\alpha_{02}=-0.5$ 
					& $-$0.036 & 0.246 & 0.246 & 95 & $-$0.024 & 0.238 & 0.243 & 95 & $-$0.035 & 0.236 & 0.242 & 96\\
					\bottomrule
				\end{tabular}
				\begin{tablenotes}
					\small
					\item \emph{Note}: SE, standard error of the parameter estimator; SEE, mean of the standard error estimator; CP, coverage percentage of the 95\% confidence interval based on the variance estimator in Section \ref{sec:infer}. Each average is based on 1000 replicates.
				\end{tablenotes}
	\end{threeparttable}}}
	\label{tab:alpha_sim}
\end{table}

\begin{figure}[!ht]
	\centering
	\includegraphics[width=\textwidth]{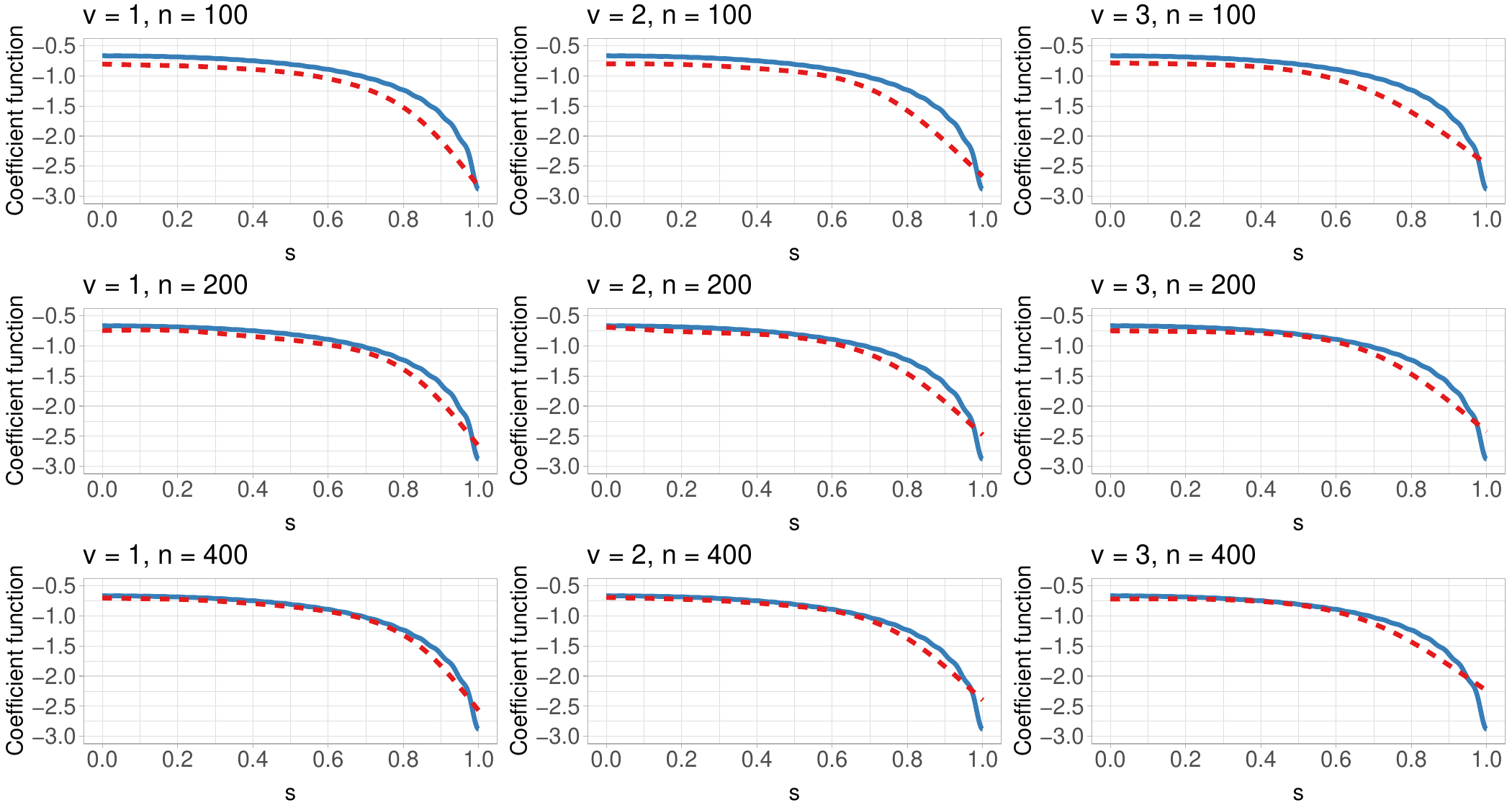}
	\caption{Estimation of the coefficient function $\beta(s)$; the solid and dashed curves show the true values and averaged estimates, respectively, where each average is based on 1000 replicates.}
	\label{fig:beta_sim}
\end{figure}

\begin{figure}[!ht]
	\includegraphics[width=\textwidth]{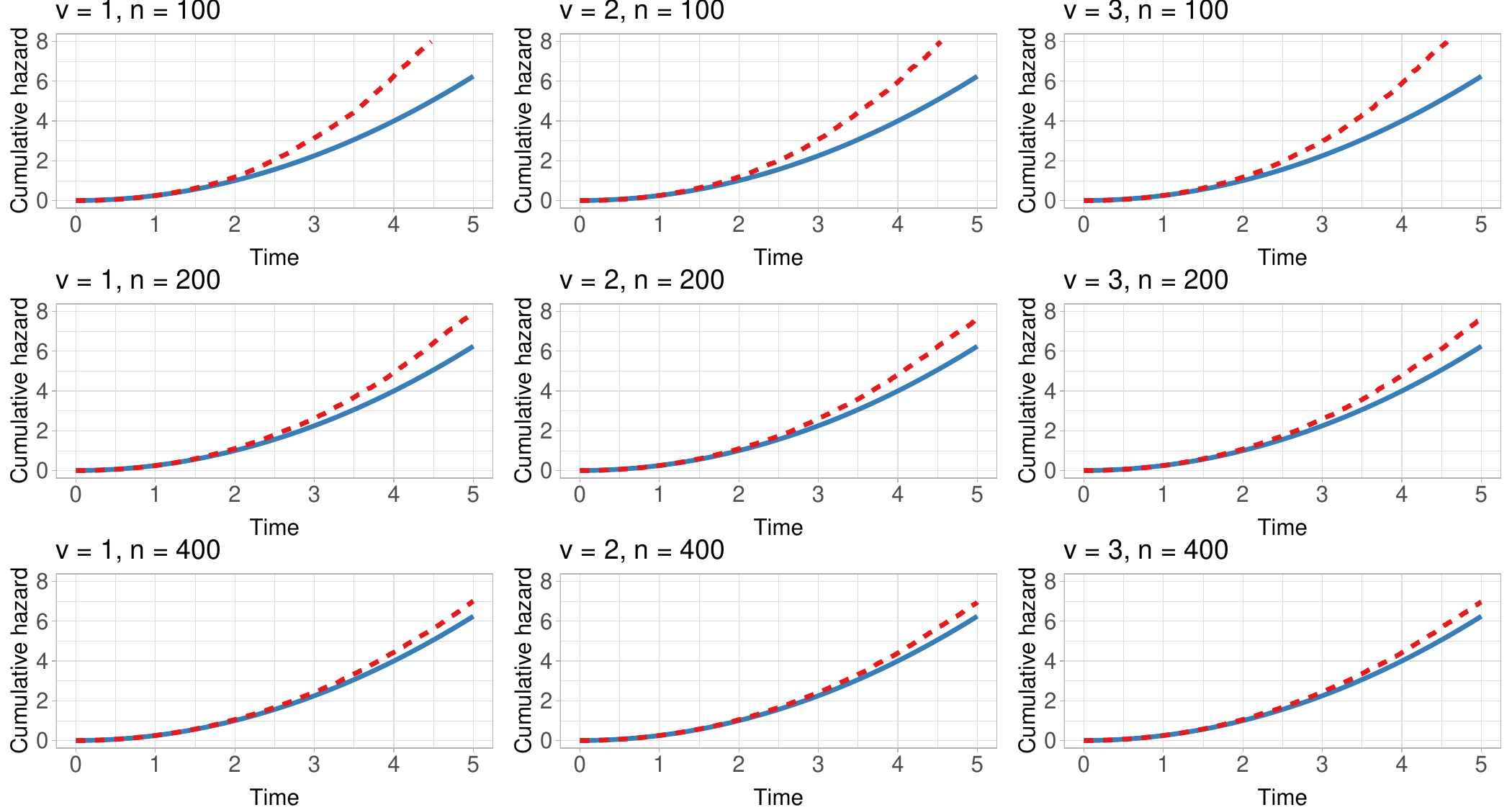}
	\caption{Estimation of the cumulative baseline hazard function $\Lambda(t)$; the solid and dashed curves show the true values and averaged estimates, respectively, where each average is based on 1000 replicates.}
	\label{fig:Lambda_sim}
\end{figure}

Figure \ref{fig:beta_sim} illustrates the performance of the estimator for the coefficient function. Under various smoothness settings of the functional covariate $Z(\cdot)$ controlled by $v$, the mean estimated curves closely align with the true coefficient functions, and the estimation accuracy improves as $n$ increases. Figure \ref{fig:Lambda_sim} demonstrates that the estimated cumulative baseline hazard function successfully recovers the true function. Although minor deviations exist at the right tail for a small sample size due to the scarcity of late events, the estimated curves converge tightly to the true curves as $n$ increases.

\begin{table}[!ht]
	\centering
	\caption{Empirical rejection rates for testing $H_{0}:\beta(s)=0$ at the nominal significance level of 0.05.}
	{\begin{threeparttable}
			\begin{tabular}{cccccccc}
				\toprule
				Sample size & Decay rate & $\omega=0$ & $\omega=0.1$ & $\omega=0.2$  & $\omega=0.3$ & $\omega=0.4$ & $\omega=0.5$ \\[4pt]
				\midrule
				$n=100$
				& $v=1$   & 0.05 & 0.22 & 0.62 & 0.95 & 1.00 & 1.00 \\
				& $v=2$   & 0.05 & 0.24 & 0.69 & 0.95 & 1.00 & 1.00 \\
				& $v=3$   & 0.04 & 0.20 & 0.66 & 0.94 & 1.00 & 1.00 \\[4pt]
				
				$n=200$
				& $v=1$   & 0.05 & 0.41 & 0.95 & 1.00 & 1.00 & 1.00 \\
				& $v=2$   & 0.04 & 0.38 & 0.95 & 1.00 & 1.00 & 1.00 \\
				& $v=3$   & 0.06 & 0.39 & 0.94 & 1.00 & 1.00 & 1.00 \\[4pt]
				
				$n=400$
				& $v=1$   & 0.05 & 0.70 & 1.00 & 1.00 & 1.00 & 1.00 \\
				& $v=2$   & 0.04 & 0.73 & 1.00 & 1.00 & 1.00 & 1.00 \\
				& $v=3$   & 0.05 & 0.72 & 1.00 & 1.00 & 1.00 & 1.00 \\
				\bottomrule
			\end{tabular}
			\begin{tablenotes}
				\small
				\item \emph{Note}: The column $\omega=0$ corresponds to the empirical Type I error, and the columns with $\omega>0$ correspond to the empirical power.
			\end{tablenotes}
	\end{threeparttable}}
	\label{tab:rejection_rates}
\end{table}

In addition, we evaluated the performance of the proposed global Wald-type test of the null hypothesis $H_{0}: \beta(s)=0$ at the nominal significance level of 0.05. To evaluate the test performance, the true coefficient function was specified as $\beta_{0}(s)=\sum_{j=1}^{50}\omega(-1)^{j} j^{-3 / 2} \psi_j(s)$, where the parameter $\omega$ dictates the signal strength. We constructed the test statistic using the first three cosine basis functions, $\psi_{1}, \psi_{2}$, and $\psi_{3}$, as the test functions. Table \ref{tab:rejection_rates} presents the empirical rejection rates of the test under different combinations of $n$ and $v$. When $\omega=0$, the null hypothesis holds, and the empirical Type I error rates tightly fluctuate between 0.04 and 0.06 across all scenarios, demonstrating that the proposed test provides accurate size control. When $\omega>0$, the empirical rejection rates represent the power of the test. As expected, the power increases monotonically as the signal strength $\omega$ deviates further from zero. For any fixed $\omega>0$, the power improves substantially as the sample size $n$ increases. Both the size and power properties remain highly robust against the varying smoothness levels of the functional covariate controlled by $v$.

\section{Application} \label{sec:application}

We analysed data from the Alzheimer's Disease Neuroimaging Initiative \citep{adni}, a multicentre longitudinal study whose primary objective is to identify and validate clinical, imaging, genetic, and biochemical biomarkers for the early detection and tracking of Alzheimer's disease. Our analysis focused on two critical endpoints, namely the time to incident mild cognitive impairment and the time to incident Alzheimer's disease. Because the clinical diagnoses of both endpoints in the study cohort are evaluated only at intermittently scheduled follow-up visits rather than through continuous monitoring, the exact timing of disease onset is interval-censored.

\begin{table}[!ht]
	\centering
	\caption{Regression analysis of disease progression across different stages.}
	{
		\scalebox{0.85}{\begin{threeparttable}
				\begin{tabular}{ll rcc rcc}
					\toprule
					& & \multicolumn{3}{c}{Mild cognitive impairment} & \multicolumn{3}{c}{Alzheimer's disease} \\
					\cmidrule(lr){3-5} \cmidrule(lr){6-8}
					Model & Risk factor & Estimate & SE & $p$-value & Estimate & SE & $p$-value \\[4pt]
					\midrule
					Functional Cox model 
					& Age $\geq$ 75      & 0.262    & 0.168 & 0.118 & 0.117    & 0.101 & 0.246 \\
					& Male               & 0.354    & 0.173 & 0.041 & $-$0.074 & 0.108 & 0.494 \\
					& Years of education & $-$0.065 & 0.031 & 0.038 & $-$0.049 & 0.017 & 0.004 \\
					& Marital status     & 0.124    & 0.179 & 0.489 & 0.249    & 0.131 & 0.058 \\
					& ApoE4              & 0.424    & 0.131 & 0.001 & 0.744    & 0.069 & $<10^{-4}$ \\[6pt]
					\midrule
					Standard Cox model   
					& Age $\geq$ 75      & 0.500    & 0.158 & 0.002 & 0.602    & 0.096 & $<10^{-4}$ \\
					& Male               & 0.460    & 0.167 & 0.006 & 0.184    & 0.104 & 0.078 \\
					& Years of education & $-$0.063 & 0.032 & 0.048 & $-$0.060 & 0.017 & 0.001 \\
					& Marital status     & $-$0.144 & 0.178 & 0.417 & 0.297    & 0.130 & 0.022 \\
					& ApoE4              & 0.399    & 0.134 & 0.003 & 0.789    & 0.068 & $<10^{-4}$ \\
					\bottomrule
				\end{tabular}
				\begin{tablenotes}
					\small
					\item \emph{Note}: SE, standard error.
				\end{tablenotes}
	\end{threeparttable}}}
	\label{tab:application_results}
\end{table}

We related the incidence of both endpoints to five baseline scalar covariates: age dichotomized as $\geq75$ years versus $<75$ years, sex, years of education, marital status, and the number of ApoE4 alleles, alongside a functional covariate of primary interest, $Z(s)$, representing the mean cortical thickness of the cerebral cortex. Because cortical thinning is a highly sensitive biomarker for neurodegeneration, modelling this structural measure continuously rather than relying on crude regional averages preserves the brain's spatial topology and enables the precise identification of localized atrophy patterns. We anatomically mapped the cortical surface onto the continuous domain $\mathcal{S}=[0,1]$ by sequentially ordering the frontal, temporal, parietal, occipital, and insula lobes of the left hemisphere, immediately followed by the corresponding lobes of the right hemisphere.

\begin{figure}[!ht]
	\includegraphics[width=\textwidth]{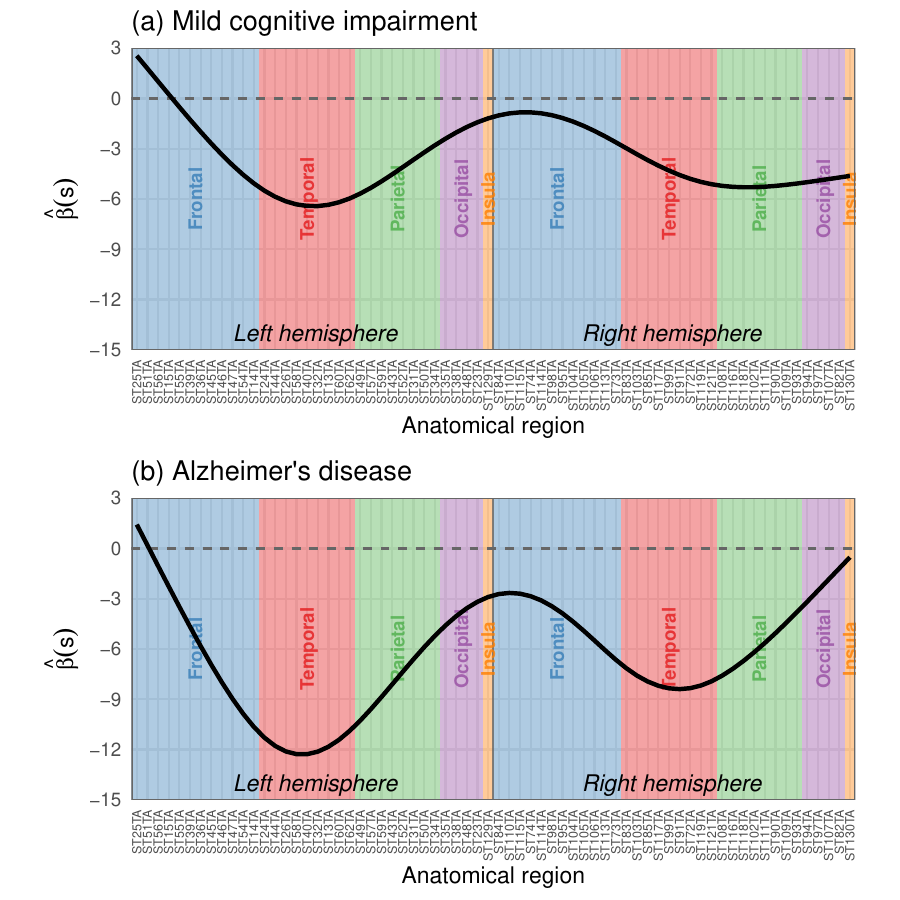}
	\caption{Estimated coefficient functions $\widehat{\beta}(s)$: (a) mild cognitive impairment; (b) Alzheimer's disease. The anatomical index $s$ maps sequentially across the left and right hemispheres, covering the frontal, temporal, parietal, occipital, and insula lobes.}
	\label{fig:beta_application}
\end{figure}

To construct the analysis datasets, we first excluded subjects with a missing baseline cognitive status, those without any follow-up visits, and those with missing covariate data. For incident mild cognitive impairment, the analysis was restricted to subjects who were cognitively normal at baseline, resulting in a final sample of 817 subjects. For incident Alzheimer's disease, the analysis was restricted to subjects who were free of Alzheimer's disease at baseline, resulting in a final sample of 2,046 subjects. We adopted the Sobolev space framework described in Example \ref{exm:sobolev} with $m=2$ and fitted model \eqref{Cox} independently to each dataset. As a comparison, we also fitted a standard Cox model \citep{Zeng2016} that adjusted solely for the scalar covariates, omitting the functional covariate entirely.

Table \ref{tab:application_results} summarizes the estimation results for the scalar covariates. The effect of age is highly significant in the standard Cox model for the incidence of both mild cognitive impairment and Alzheimer's disease, yielding $p$-values of 0.002 and below $10^{-4}$, respectively. However, this effect is completely attenuated and becomes nonsignificant in the functional Cox model, where the respective $p$-values rise to 0.118 and 0.246. These findings suggest that the impact of age is largely mediated by the anatomical patterns of cortical atrophy captured by $Z(s)$. In contrast, the number of ApoE4 alleles remains a robust and independent risk factor. The estimated effects are remarkably consistent between the two models and highly significant for both incident mild cognitive impairment and incident Alzheimer's disease, yielding $p$-values at or below 0.003. The persistence of this genetic effect implies that the associated risk operates through biological pathways beyond mere structural atrophy. Finally, more years of education consistently exhibit a protective effect against both incident mild cognitive impairment and incident Alzheimer's disease across both models.

We formally tested the global null hypothesis $H_0: \beta(s) = 0$ using the proposed Wald-type test, employing the first $L_n$ cosine basis functions as the test functions. For incident mild cognitive impairment, the test statistic constructed with $L_n=5$ was 17.819, yielding a $p$-value of 0.003. For incident Alzheimer's disease, the test statistic constructed with $L_n=8$ was 239.972, yielding a $p$-value below $10^{-5}$. These findings confirm that spatial patterns of cortical thickness have a significant effect on both endpoints. Figure \ref{fig:beta_application} displays the estimated coefficient functions $\widehat{\beta}(s)$. The estimated curves reveal that severe cortical thinning in the left temporal lobe is associated with an increased hazard of both incident mild cognitive impairment and incident Alzheimer's disease, although the magnitude of this effect is more pronounced for the latter. Specifically, for incident Alzheimer's disease, $\widehat{\beta}(s)$ reaches its global minimum at a value of approximately $-12$ in the left temporal lobe, indicating a drastically increased risk.

\section{Discussion} \label{sec:conclusion}
To characterize the effect of a functional covariate on an interval-censored survival outcome while adjusting for scalar covariates, we study the functional Cox model within a reproducing kernel Hilbert space framework. Computationally, we establish a representer theorem to devise a highly stable EM algorithm for model fitting, accompanied by a novel scheme for tuning parameter selection. Theoretically, we derive the convergence rates of the proposed estimators and prove the asymptotic normality of the estimators for both the regression parameters and linear functionals of the coefficient function. Furthermore, we show that their asymptotic variances attain the semiparametric efficiency bounds. These limiting distributions provide a rigorous foundation for developing formal inferential tools to quantify estimation uncertainty.

Although the proposed global test for $\beta(\cdot)$ evaluates the overall significance of the functional covariate, it does not provide formal inference to assess the statistical significance of localized regional effects. A natural solution is to construct a simultaneous confidence band for $\beta(\cdot)$, which accounts for multiple comparisons across the continuous domain $\mathcal{S}$ and reliably identifies specific brain regions driving disease progression. Accomplishing this task for the functional Cox model with interval-censored data presents unique theoretical challenges. The fundamental difficulty lies in establishing the weak convergence of $\widehat{\beta}(\cdot)$ in the $l^{\infty}(\mathcal{S})$ space, which is severely complicated by the slow-converging estimator $\widehat{\Lambda}$.

Our current analysis treats incident mild cognitive impairment and Alzheimer's disease as separate endpoints, despite their sequential nature in neurodegeneration. Thus, extending the proposed model to handle such interval-censored multistate data in future research would enable a joint analysis of the entire disease trajectory. By properly addressing the multiple transition times under interval-censoring, such joint analysis could enhance the statistical efficiency and clinical relevance of findings derived from longitudinal studies like the Alzheimer's Disease Neuroimaging Initiative.









\bibliographystyle{apalike}
\bibliography{paper-ref}

\end{document}